\documentclass[aps,prl,twocolumn,showpacs,superscriptaddress,floatfix,
amsmath,amssymb]{revtex4}
\usepackage{graphicx}
\begin{document}
\title{Duality Between the Weak and Strong
Interaction Limits for Randomly Interacting Fermions}
\author{Philippe Jacquod}
\affiliation{Instituut-Lorentz, Universiteit Leiden
P.O. Box 9506, 2300 RA Leiden, The Netherlands}
\author{Imre Varga}
\affiliation{Elm\'eleti Fizika Tansz\'ek, 
Budapesti M\H uszaki \'es Gazdas\'agtudom\'anyi Egyetem,
H-1521 Budapest, Hungary}
\affiliation{Fachbereich Physik, Philipps--Universit\"at Marburg,
Renthof 6, D-35032 Marburg, Germany}
\date{\today}
\begin{abstract}
We establish the existence of a duality transformation for generic
models of interacting fermions with two-body interactions. The eigenstates
at weak and strong interaction $U$ possess similar statistical
properties when expressed in the $U=0$ and $U=\infty$ eigenstates bases 
respectively. This implies the existence of a duality point $U_d$ where 
the eigenstates have the same spreading in both bases.
$U_d$ is surrounded by an interval of finite width which is characterized
by a non Lorentzian spreading
of the strength function in both bases. Scaling arguments predict
the survival of this intermediate regime as the number of particles is
increased.
\end{abstract}
\maketitle{}

In noninteracting systems, many-body fermionic states are
totally antisymmetrized products of one-body states, so-called
Slater determinants. For strongly interacting
fermions on the other hand, the eigenstates 
expressed in the basis of Slater determinants have a large number of 
nonzero components, since then the one-orbital occupation is in general
no longer a good quantum number. As the 
interaction is turned on, the number of such components increases,
and this crossover from the weak to the strong interaction regime
has been investigated both from the point of view of the 
statistical properties of the spectrum and the eigenstates
\cite{aberg90,kusnezov96,berkovits99,waintal99,song00,sahu01}.
This crossover determines the threshold in excitation energy
above which Dyson's random matrix
theory applies for atomic levels (as e.g. in the Ce atom) \cite{flambaum94} 
or nuclear energy levels (as obtained in neutron scattering 
experiments) \cite{zelev} and below which, for example, 
pine
quasi-particle excitations in 
chaotic quantum dots are well defined \cite{altshuler97}.

In this article we investigate the reversed crossover from strong to weak 
interaction. Relying both on analytical arguments and exact numerical 
diagonalization of 
systems with few particles, we show that the spreading of the 
eigenstates over the strong interaction basis as the interaction is 
decreased, is similar to the spreading over the weak noninteracting 
basis as the interaction is increased. This results in a
duality transformation between the strong and weak interaction
limits. The fixed (dual) point of this transformation
lies inside a finite-width intermediate regime,
which is characterized by a non-Lorentzian spreading and a
maximal complexity (which we define below through the structural entropy) 
of the eigenstates in both bases. 
Scaling arguments predict the survival of this intermediate regime
as the number of particles increases.
In this regime, eigenstates 
properties cannot be extracted perturbatively from any of the asymptotic 
limits, but are well captured by our method. The eigenstate structure 
suggests the
existence of quasi-particle excitations in the strong interaction
limit, which we were, however, unable to determine.

We consider the deformed Two-Body Random Ensemble (TBRE) 
for $n$ interacting spinless fermions \cite{french71,brody81,kota} 
\begin{eqnarray}\label{hamiltonian}
H & = & H_0+U H_1\nonumber \\ & = & \Lambda\bigl(\sum \epsilon_{\alpha}
c^{\dagger}_{\alpha} c_{\alpha}
+ U \sum
  Q_{\alpha,\beta}^{\gamma,\delta}
c^{\dagger}_{\alpha} c^{\dagger}_{\beta} c_{\delta}
c_{\gamma}\bigr).
\end{eqnarray}
\noindent The $m$ different one-body energies are distributed as
$\epsilon_{\alpha} \in [-m/2; m/2]$ so as to fix the mean level spacing 
$\Delta \equiv 1$. The interaction matrix elements 
$Q_{\alpha,\beta}^{\gamma,\delta}$ are
independent random variables with a zero-centered Gaussian distribution of
unit variance. The parameter
$\Lambda=\Delta/(U+\Delta)$ has been introduced to keep
the density of states roughly constant as $U$ varies.
In the $U=0$ eigenstate basis, the Hamiltonian is represented 
by a $N \times N$ matrix of size $N=(^m_n)$ 
with $K=1+n(m-n)+n(n-1)(m-n)(m-n-1)/4$ nonzero
matrix elements per row. For a sufficiently large 
number of particles, the many-body Density of States (DOS)
is well approximated by 
a Gaussian of width $B_n \sim \sqrt{K} \Lambda$ \cite{brody81,kota}. Below 
we will investigate the properties of levels in the middle of 
the DOS.

In the $U=0$ basis (superscripts 
$(0)$ and $(\infty)$ indicate the corresponding basis) 
the eigenstates structure is conveniently
described by the strength function (SF)
\begin{equation}
\rho^{(0)}(E) = \bigl\langle \sum_{A} |\psi_{A}(I)|^2 
\delta(E+E_I-E_{A}) \bigr\rangle_I ,
\end{equation}
\noindent and the inverse moments
\begin{equation}
\xi^{(0)}_p = \bigl\langle \bigl(\sum_{I}
|\psi_{A}(I)|^{2p}\bigr)^{-1} \bigr\rangle_A,
\end{equation}
\noindent where the averages $\langle ... \rangle_{I,A}$ are taken 
either over the
eigenstates $\phi_{I}$ (with eigenvalues $E_{I}$)
of $H_0$ or the eigenstates $\psi_{A}$ (with eigenvalues $E_{A}$) of $H$. 
The moment $p=2$ is the participation 
ratio (PR) and gives the typical number
of nonzero components $\psi_{A}(I)$. 
For model (1) one has $\xi^{(0)}_p \in 
[1,\xi^{\rm GOE}_p = N^{p-1}/(2p-1)!!]$, where the maximal value
corresponding to Dyson's Gaussian Orthogonal Ensemble (GOE) 
is achieved for $U=\infty$ \cite{kaplan00}.
There are, beside the one-body spacing $\Delta$, two important 
energy scales \cite{aberg90} : 
the mean spacing between states directly 
coupled by the two-body interaction $\Delta_c^{(0)} = B_2^{(0)}/K 
\approx 4\Delta/mn^2$ and the $n$-body spacing 
$\Delta_n^{(0)} = B_n^{(0)}/N$, where $B_n^{(0)} \approx n(m-n)\Delta$ 
is the $n$-body band (and hence $B_2^{(0)} \approx 2(m-2)\Delta$). 
As $U$ increases, it has been found
that $\rho^{(0)}(E)$ undergoes two transitions from a delta peak to a
Lorentzian shape first, 
then to a Gaussian shape \cite{aberg90,kota,flambaum97}. In the Lorentzian
regime, the width $\Gamma^{(0)}$ of $\rho^{(0)}(E)$ is given by
the golden rule $\Gamma^{(0)} \propto U^2/\Delta_c^{(0)}$ and the
PR is $\xi^{(0)}_2 \approx \Gamma^{(0)}/\Delta_n^{(0)}$
\cite{aberg90}. The Lorentzian regime is defined
by the two conditions $\xi^{(0)}_2 \gg 1$ and 
$\Gamma^{(0)} < B^{(0)}_n$ \cite{flambaum97,jacquod95}. In the dilute
limit $1 \ll n \ll m$, these conditions translate into
\begin{equation}\label{u0lorentz}
\Delta/\sqrt{n N} \ll U < \Delta/\sqrt{n}.
\end{equation}
\noindent When the Gaussian regime is
entered, the SF spreads over the full bandwidth so that 
the eigenstates have a finite fraction of nonzero components
$\xi^{(0)}_2 = O(N)$.
\begin{figure}
\includegraphics[width=3.4in]{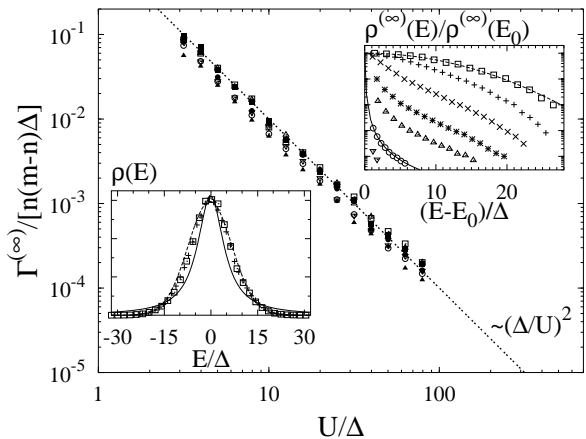}
\caption{\label{fig:fig1}
Width of the strength function $\rho^{(\infty)}$ as a function
of $U/\Delta$ for $n \in [3,6]$ and $m \in [10,14]$. The dotted line
indicates the Golden rule predicted behavior (\ref{gruleinf}). 
Upper inset: crossover of $\rho^{(\infty)}$ from delta peak to Lorentzian
to Gaussian shape for $m=12$, $n=5$ and $U/\Delta = $
0.55 ($\square$), 1.09 ($+$), 2.15 ($\times$), 4.26 ($\ast$), 8.43
($\triangle$), 33 ($\bigcirc$), and 129.15 ($\triangledown$).
The solid and dashed lines give a Lorentzian (solid) and a Gaussian
(dashed) fit 
respectively. Lower inset: strength functions $\rho^{(\infty)}$ 
($\square$) and $\rho^{(0)}$ ($+$) at the dual point $U_d \approx 1.11$ 
for $m=12$ and $n=5$. Note that the shape is neither Lorentzian (solid line) 
nor Gaussian (dashed line).}
\end{figure}
Having summarized some of the known results for the eigenstates
structure as the interaction is switched on, we now turn our attention
to the reversed problem and decrease $U$ starting from $U=\infty$.
Both the SF $\rho^{(\infty)}$
and the moments $\xi^{(\infty)}_p$ can be defined in the same way as
above provided the $U=\infty$ basis is
fixed individually for each realization of the Hamiltonian (1).
Since the occupation operator $n_{\alpha}=c^{\dagger}_{\alpha} c_{\alpha}$
does not commute with $H_1$, $H_0$ induces one-body transitions between 
different eigenstates of $H_1$, leading to an increase of 
$\xi^{(\infty)}_p>1$ and a broadening of $\rho^{(\infty)}$
as $U$ decreases. The $H_0$-induced
transitions also lead to two successive crossovers of the SF
$\rho^{(\infty)}$, first from a delta peak to a Lorentzian shape, 
then to a Gaussian shape, and this 
is shown on the upper inset to Fig.~\ref{fig:fig1}.
As is the case for $\Gamma^{(0)}$,
the Golden rule gives a good estimate of the width $\Gamma^{(\infty)}$
of the SF in the Lorentzian regime, as we now proceed to show. 
Under the assumption that for $U=\infty$, the TBRE has
random ergodic eigenstates in the middle of its spectrum \cite{kaplan00}, 
all eigenstates are directly connected to each other by $H_0$ so that 
$\Delta_c^{(\infty)}=\Delta_n^{(\infty)}=B_n^{(\infty)}/N$.
In the limit of large number of particles and orbitals,
the $U=\infty$ density of states is well approximated by
a Gaussian with a width 
$B_n^{(\infty)} \approx \sqrt{K} \Lambda U \approx n(m-n)  \Delta$
\cite{brody81,kota}, and one has $\Delta_n^{(\infty)} \approx \Delta_n^{(0)}
\propto n (m-n) \Delta/N$. Assuming random ergodic many-body 
eigenfunctions at $U=\infty$ \cite{brody81,kota}, the transitions 
matrix elements have a variance
$\langle H_0^2 \rangle = \sum_{\alpha} (\epsilon_\alpha/U)^2/N
\approx (n (m-n) \Delta^2/U)^2/N$. 
The Golden rule then predicts 
\begin{equation}\label{gruleinf}
\Gamma^{(\infty)} \propto n (m-n) \Delta^3/U^2.
\end{equation}
\noindent This prediction is confirmed by numerical data presented in 
Fig.~\ref{fig:fig1}.
These data, as well as those to be presented below, have been 
obtained via exact diagonalization
of systems of up to $N=3432$ (corresponding to $n=7$ and $m=14$),
performing averages over 20 (for the largest
systems) to 500 realizations of Hamiltonian (1) for each parameter set
and avoiding the tails of the DOS by
keeping only 33\% of the states in the middle of the spectrum.

\begin{figure}
\includegraphics[width=3.4in]{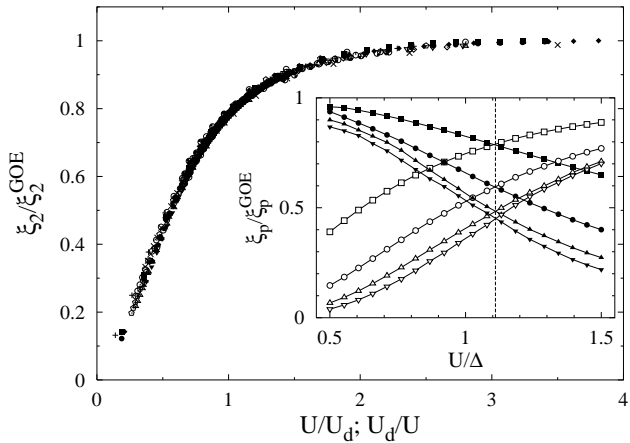}
\caption{\label{fig:fig2}
$\xi_2^{(0)}/\xi_2^{\rm GOE}$ (empty symbols) and 
$\xi_2^{(\infty)}/\xi_2^{\rm GOE}$ (full symbols)
vs. the
rescaled interactions $U/U_d$ and $U_d/U$ respectively. Different symbols
correspond to different $n \in [2,7]$ and $m \in [10,16]$.
Inset: $U$-dependence of $\xi_p^{(0)}$ (empty symbols)
and $\xi_p^{(\infty)}$ (full symbols) for $p=2$ ($\square$),
3 ($\bigcirc$), 4 ($\triangle$) and 5 ($\triangledown$), for 
$m=12$ and $n=5$. The dashed line indicates
the ($p$-independent) intersection point $U_d/\Delta \approx 1.11$.}
\end{figure}
In the Golden rule regime, the PR is
given by $\xi_2^{(\infty)} = 
\Gamma^{(\infty)}/\Delta_n^{(\infty)} \approx N (\Delta/U)^2$
and according to the conditions $\xi_2^{(\infty)} > 1$
and $\Gamma^{(\infty)} < B_n^{(\infty)}$, the
Lorentzian regime in the $U=\infty$ basis is bounded by the 
inequalities
\begin{equation}\label{uinflorentz}
\Delta < U < \sqrt{N} \Delta.
\end{equation}
\noindent This is confirmed qualitatively 
by the upper inset to Fig.~\ref{fig:fig1} and by
numerical data to be published elsewhere \cite{imre02}.

The remarkable fact that $\Delta_n^{(\infty)}$ and
$\Delta_n^{(0)}$ have the same parametric dependence in $n$
and $m$ implies that $\Gamma^{(0)}=\Gamma^{(\infty)}$
and $\xi_2^{(0)}=\xi_2^{(\infty)}$ can be both satisfied 
for $U_d \sim \Delta/n^{1/4}$. This is
illustrated on the lower inset to Fig.~\ref{fig:fig1} and the inset to 
Fig.~\ref{fig:fig2}. 
Remarkably enough, we have found that at $U_d$, the PR's
take a universal value $\xi_2^{(0)}=\xi_2^{(\infty)} 
\approx 0.8 ~\xi_2^{\rm GOE}$ independently on 
$n$ and $m$. This calls for a rescaling 
$U \rightarrow U/U_d$ for $\xi_2^{(0)}/\xi_2^{\rm GOE}$ and
$U \rightarrow U_d/U$ for $\xi_2^{(\infty)}/\xi_2^{\rm GOE}$
after which all data for the PR 
fall on top of each other, as can be seen on Fig.~\ref{fig:fig2}.

\begin{figure}
\includegraphics[width=3.4in]{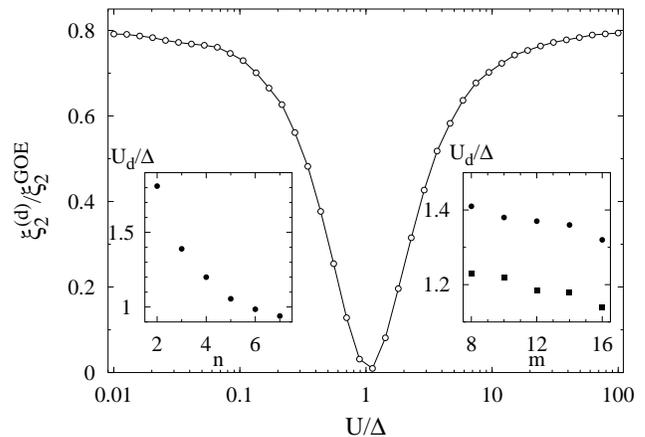}
\caption{\label{fig:fig3}
$\xi_2^{(d)}/\xi_2^{\rm GOE}$ calculated in the 
$U=U_d$ basis for $m=12$ and $n=5$. Left inset: $n$-dependence
of $U_d$ for $m=14$. Right inset: $m$-dependence of $U_d$
for $n=3$ ($\bigcirc$) and 4 ($\square$).}
\end{figure}

The inset to Fig.~\ref{fig:fig2} shows that at $U_d$, the higher 
moments $\xi_p$ with
$p>2$ also cross. This behavior has been found to hold for all
$m$ and $n$. It implies that the
eigenstates have exactly the same spreading over both bases, 
and that the SF's have not only the same width, but also exactly the same
shape at $U_d$ ($\xi_p^{-1}$ are moments of the SF), as shown in 
the lower inset to Fig.\ref{fig:fig1}. Moreover, we found
that at the dual point the SF satisfies the scaling
$\Gamma(U_d)\propto m n^{3/2}\Delta$ as follows from (\ref{gruleinf})
and $U_d\propto\Delta /n^{1/4}$.
We then calculate the PR $\xi_2^{(d)}$
in the $U=U_d$ (dual) basis, and numerical results are shown in 
Fig.~\ref{fig:fig3}. The 
symmetry around $U=U_d$ is evident, and for both $U \rightarrow 0$ and 
$U \rightarrow \infty $ limits, there is saturation at $\xi_2^{(d)} 
\approx 0.8 ~\xi_2^{\rm GOE}$. These results establish the existence of
a duality transformation $U^{*}=U_d^2/U$
connecting the strong and weak interaction regimes, so that the PR's 
and the SF satisfy $\xi_{2}^{(0)}(U) = \xi_{2}^{(\infty)}(U^{*})$,
$\xi_{2}^{(d)}(U) = \xi_{2}^{(d)}(U^{*})$,
$\rho^{(0)}(U) = \rho^{(\infty)}(U^{*})$ and 
$\rho^{(d)}(U) = \rho^{(d)}(U^{*})$. 

The above estimate for $U_d \sim \Delta/n^{1/4}$ relies on 
the assumptions that the Golden rule correctly estimates the
width of the SF in both bases, and that $\xi_2^{(0,\infty)} = 
\Gamma^{(0,\infty)}/\Delta_n^{(0,\infty)}$.
Equivalently, this requires to be in the Lorentzian regime,
which is however impossible as the
two conditions (\ref{u0lorentz}) and (\ref{uinflorentz}) 
are mutually exclusive, and the predicted $U_d \sim \Delta/n^{1/4}$ lies 
outside both Lorentzian regimes. One may thus wonder if
its predicted parametric dependence in $m$ and $n$ makes any sense
at all. $U_d$ as a function of both $m$ and $n$ is shown
in the insets to Fig.~\ref{fig:fig3}. First, it is seen that the 
$m$-dependence of
$U_d$ is very weak (we found $U_d \sim \Delta m^{-\alpha}$ with $\alpha 
<0.08$ for a range $m \in [8,20]$). Even though quite weak,
this $m$-dependence presumably indicates that we
are not deep enough in the dilute limit. Second we extracted 
the $n$-dependence of $U_d$ for $m=14$ where we had the largest
range $n \in[2,7]$. We got $U_d \sim \Delta n^{-\beta}$, with an 
exponent $\beta \in [0.3;0.5]$. These bounds
on $\beta$ are compatible with the Golden rule estimates which 
exclude $U_d$ from both Lorentzian regimes, i.e. 
\begin{equation}\label{inter}
\Delta/\sqrt{n} < U_d < \Delta,
\end{equation}
giving $\beta \in [0,0.5]$.
We conclude that $U_d \sim \Delta / n^{1/4}$ is in good qualitative 
agreement with our numerical results, and that
the inequalities (\ref{inter}) define an intermediate regime which is 
nonperturbative in the two bases, and whose
width increases parametrically with $n$. 

\begin{figure}
\includegraphics[width=3.4in]{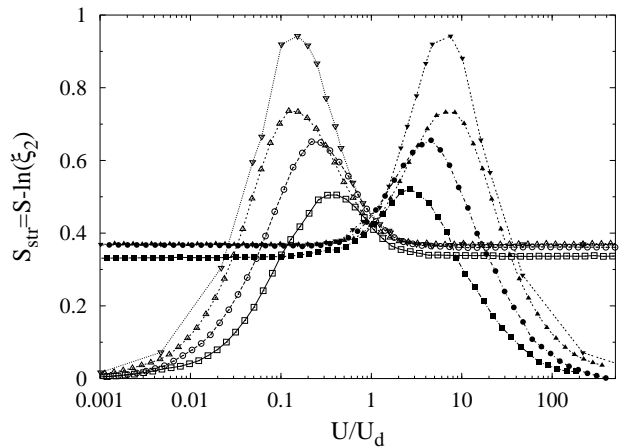}
\caption{\label{fig:fig4}
Structural entropy $S_{str}=S- \ln(\xi_2)$ in the $U=0$ (empty
symbols) and $U=\infty$ (full symbols) basis
for $m=14$, and $n=2$ ($\square$), 3 ($\bigcirc$), 4 ($\triangle$),
5 ($\triangledown $).}
\end{figure}
Additional structures in the wave functions can be captured by the
structural entropy $S_{\rm str}$, 
which is defined by subtracting from the Shannon 
entropy $S=\langle\sum_{I}|\psi_A(I)|^2\ln 
|\psi_A(I)|^2\rangle_A$ the log of the PR:
$S^{(0,\infty)}_{\rm str}=S^{(0,\infty)}-\ln\xi^{(0,\infty)}_{2}$ 
\cite{pipek91}. It measures the contribution to the Shannon 
entropy which is not contained in the PR, and thus not in the bulk of the 
SF. Numerical results are 
presented in Fig.~\ref{fig:fig4}. Expectedly, the crossing 
of $S^{(0,\infty)}_{\rm str}$ 
occurs at $U_{d}$ (since all the moments $\xi_p$ determine $S_{\rm str}$)
but the crucial feature of Fig.~\ref{fig:fig4} is that almost immediately
after the crossing, $S_{\rm str}$ takes on its asymptotic GOE 
value $S_{\rm str}^{\rm GOE}=0.3689\ldots$  - and therefore $U_{d}$ is the 
crossover point to the regime with GOE--like behavior
{\it in both bases}. The second crucial feature of Fig.~\ref{fig:fig4} 
are the peaks in $S_{\rm str}$ located on both sides of
$U_d$. Such peaks have already been observed in various models
where they
indicate unusually large fluctuations of eigenstates \cite{power}.
It is expectable that these peaks extend over the intermediate, 
non-Lorentzian regime $\Delta/\sqrt{n} < U < \Delta$, which is
borne out by the data plotted in Fig. 4, where
once $S_{\rm str}$ is plotted against
$U/U_d \simeq U n^{1/4}/\Delta$, the peaks broaden symmetrically as more 
particles are added.

Examples of models where such a duality between localized 
(with $\xi_{2}=1$) and delocalized (with $\xi_{2}=O(N)$) asymptotic 
regimes is related to a sharp metal-insulator transitions are
provided by one-dimensional lattices with quasi-periodic potential 
(where the duality connects momentum and spatial eigenfunction 
coordinates) \cite{aubry79} and tight-binding models for strongly
correlated fermions \cite{florian}. Our model is however
fundamentally different in that
the duality point $U_d$ is protected by a finite-sized
interval characterized by a non-Lorentzian spreading of the 
eigenstates over a finite fraction of both the strong and weak interaction 
eigenstate. According to (\ref{inter}), this intermediate regime 
survives as $m$ and $n$
increase and this results in a smooth crossover (and not a sharp
transition) to fully
developed chaos, characterized by the collapse of all the data points
over one single curve shown on Fig.~\ref{fig:fig2} (and not a 
single crossing of the
curves at $U/U_{d}=1$). This is to be put in perspective with 
the recent controversy over
whether or not particle-hole
excitations undergo a delocalization transition in Fock space
similar to the Anderson transition as their excitation energy increases
\cite{aberg90,altshuler97,song00} (increasing the 
excitation energy reduces the level spacing and is thus equivalent to 
increasing the interaction strength $U$). The results we presented 
corroborate the conclusions drawn in \cite{aberg90} of the existence of
a smooth crossover and not a sharp transition.

In summary we have established the existence of a duality 
transformation between the weak and strong interaction regimes
of deformed TBRE. At the duality point $U_d$, the eigenstates
have GOE fluctuations (for $U >$ $(<) U_d$
in the $U=0$ $(\infty)$ basis) and $U_d$ is surrounded by a finite-sized, 
intermediate regime where both limiting bases fail to describe the
eigenstates. Together with the Lorentzian form of 
$\rho^{(\infty)}$ at large $U \gg \Delta$, this duality suggests the
existence of quasiparticle excitations at $U_d \ll U \le \infty$, which 
we were however not able to identify.
The existence of this duality transformation is 
related to the two-body nature of the interaction. For $k$-body 
interaction, $n \gg k>2$, the $U= \infty$ bandwidth is given by
$B_{n}^{(\infty )} \sim \Delta n^{k/2}m^{k/2} > B_{n}^{(0)}$ 
\cite{brody81,kota}, and the PR in both bases are equal to each other 
for $U_d \propto \Delta m^{1-k/2}n^{3/4-k/2}$.
We then have $\Gamma^{(0)}(U_d) \propto \Delta m n^{3/2}$
while $\Gamma^{(\infty)}(U_d) \propto \Delta m^{k/2} n^{1/2+k/2}$
so that the SF's
should differ at the intersection point of the PR's for $k \ne 2$.

This work was supported by the Swiss National Science Foundation, the
Dutch Science Foundation NWO/FOM, the U.S. Army Research Office,
the Alexander 
von Humboldt Foundation, the Hungarian Research Fund (OTKA) under T029813, 
T032116 and T034832 the Education Ministry of Hungary under
MEC-00238/2001 and by the European Community's Human Potential
Programme under contract HPRN-CT-2000-00144, Nanoscale Dynamics.

\end{document}